# An approach to standardize, automate omni-channel and AI transactional digital service creation

**Antoine Aamarcha, Martin Caussanel, Hadrien Lanneau**
**Kevin Mege, Florian Peyron**
Prisme.ai, France
{antoine.aamarcha,martin.caussanel,hadrien.lanneau,kevin.mege,hadrien.lanneau}@prisme.ai

## Abstract

Our work is at the crossroads of two categories of technologies. On the one hand, omnichannel digit services, to address the needs of users in the most seamless way. On the other hand, low code approaches, to build simply even complex software applications. In this twofold context, we propose DSUL (Digital Service Universal Language). It allows to build omnichannel services with minimal work from their designers. We describe precisely how DSUL operates, and its innovation in regard to the state of the art. We also consider the various methods to evaluate this framework.

## 1 Introduction

In a decade, Omnichannel digital services have become the main trend for online activities. They offer several advantages to the user, who can interact with the services through various channels: chat, native web interactions (form, table, list …), smartphones, vocal assistants, etc. these channels being synchronized. Consequently, these services are adapted to new, and often complementary, usages: messaging channels, social networks, vocal assistants, besides classical Web interaction. They also allow the user to choose the most relevant channel for his situation, wherever he is: at the office working on applications on his browser, on public transport looking for the next stop on his smartphone, at home cooking while asking by voice the next step in his recipe, in the car, hands full, asking for the nearest power station, in the metaverse asking for help from a virtual avatar, etc. Thus, omni-channel services are very convenient for users.

But the conception and implementation of these services usually has a high cost. Indeed, each channel has its own characteristics; they require services to be adapted. Besides, the coordination of these various services also requires additional work.

This workload also has environmental consequences. Indeed, the corresponding code, and its handling (hosting, execution, CI/CD pipelines, etc.) require resources, notably energy. In a context of sustainability[1], lighter solutions are necessary.

In answer to these issues, we propose a new paradigm to reduce, by several orders of magnitude, the resources and time necessary to build an omnichannel digital service. Besides, our solution allows the standardization, the automation, and high reusability of the services.
In a first time, we describe the current state of the art regarding conception of omnichannel services. We then present in detail our approach and its advantages. Finally we propose the main leads for its evaluation.

## 2 Related Work

There are many approaches to facilitate the development of digital services for several channels. We propose here a quick overview of the main ones, for a better positioning of our approach.

---

[1] Data Centers consume 1% of the global energy use (https://www.iea.org/reports/data-centres-and-data-transmission-networks; IEA is the International Energy Agency)

Most of these solutions belong to the low-code trend: remind that low code platforms are defined, in the seminal report (Richardson & Rymer, 2014, p. 3), "as Platforms that enable rapid application delivery with a minimum of hand-coding, and quick setup and deployment, for systems of engagement.". However, other works relate to our, without using low code strictly speaking.

Thus, several research-oriented approaches are related to chatbot development, since chatbot is a very large trend in commercial services. Several papers are based on AIML[2] (Artificial Intelligence Modeling Language); for instance (AbuShawar *et al.* 2015) or (Satu *et al.* 2015). From these papers, it appears that AIML is functional for simple dialog use cases. However, they fail to address more complex or diverse use cases, for several reasons:

- Dialog transaction features: slot-filling, retry, etc. are not supported;

- They do not allow coding structure controls: conditions, event triggers, loops …

- Omnichannel development (as defined above) is not possible;

- Extension to other channels or other systems, for instance with addition of one's own code, is not allowed.

(Tan & Inamura 2012) addresses some of these issues. Their framework takes into account the context of the dialog, and multimodality. Nonetheless, they do not enable omnichannel services, dialog transaction features, or structure controls.

(Daniel *et al.* 2020) proposes a more complete framework: it manages all dialogue features, and takes into account context and transaction events. Services on voice/written chat channels are possible, but not on web-based channels. From a software perspective, control structures are not taken into account.

There are also solutions to ease coding related to Machine Learning (ML). For instance, Maiya (2020) aims to simplify all steps in a classical ML workflow: data preprocessing, training, tests, etc. But this work focuses on ML applications only.

On the market, numerous frameworks are available, such as Lowdefy (https://lowdefy.com), Mendix (www.mendix.com) or Apian (https://appian.com) for instance. But each of these frameworks focus on one or few single dimensions of software coding, for instance UI, CRM, or the automation of processes. This specialization of low-code frameworks is formalized in (Richardson & Rymer 2016). This report defines 5 categories: database, request-handling, mobile first, process, and general purpose. The solutions belonging to this last category are more general than for the other ones. Still, DSUL offers more features than them, as we explain in the next section.

Our review shows that each of the existing solutions simplify only one or a little number of dimensions of development of digital services, but not all. It could be appealing to combine these various tools; but doing so would then require a lot of coordination work, if ever possible.

Our proposition can address at once all dimensions. We present it in the next section.

## 3 Description of Digital Service Universal Language

Our solution is primarily used to build Omnichannel digital services. Examples of such services are Websites, chatbots, digital forms, smartphone applications, etc. The solution proposed allows to generate all services for an application (for instance: chatbot and form) from a single, simple yaml file, defined by Digital Service Universal Language (DSUL).

DSUL provides an abstraction which allows to describe all components, events, and actions of a given service. It also handles the processes carried out by components: actions, Model training (for Machine Learning), etc. The service defined by DSUL corresponds to an oriented graph.

More precisely, DSUL meet several requirements:

- Easiness of coding: our framework provides a significant saving in the time and the complexity to develop the components. Part of the work of creation of services can be automated;

---

[2] http://www.aiml.foundation

- Omnichannel uses: DSUL allows to develop services accessible from any targeted channel; besides, it proposes natively to integrate all channels for a given service;

- Extended coding possibilities, notably for structure controls;

- Openness to other systems: thanks to the possibility to wrap any customized code in DSUL;

- Possible use for Machine Learning purposes: accelerated evaluation of models and algorithms, deployment, training. This requirement also covers scientific uses, for instance quick prototyping for evaluations;

- Standardization: with all the features described above, DSUL is aimed to be a standard for creating any omnichannel digital service with one single language. Standardization is important in computer application, since it allows to focus on a single framework and support interoperability;

- Green coding considerations: the development easiness, in turn, entails a lesser consumption of resources needed to handle the digital service.

Most of these requirements are addressed by state-of-the-art solutions, as our study shows. However, each of these solutions meets only a few requirements. On the contrary, our approach proposes an integrated answer.

Our approach has some similarities to low-code concepts, but differs because at the core of DSUL lies the language and its properties (syntax, etc.). A WYSIWYG editor provides an even easier edition of the file, notably with the visual representation of the service graph; but this editor is an additional layer.

The complementary piece of DSUL is the Runtime. It acts as an interpreter which parses the file, and carries out the execution of its instruction.

Throughout this section, we describe the main features of DSUL, and their contribution to the various target requirements mentioned above.

## 3.1 Framework Overview

Before describing the details of our Language, we provide a first set of definitions for the concepts we used (NB: all the documentation is available at https://docs.eda.prisme.ai/en/workspaces/). Then we present the software architecture, and finally some examples.

### 3.1.1 Definitions

A digital service includes a Frontend (what the end user can see and interact with) and the Backend (the services which handle inputs and outputs from/to the end user, without being visible).
In this context, let us consider the main software components used to build a service:

- Frontend:

  o Workspace: it is a project, in which all other components are defined. A workspace can be published as an App (more details below), and then used in other workspaces. This feature provides several advantages:

    - Facilitate the cooperation between transversal contributors, for instance developers and non-developers: each one handle the workspaces which suit to his specialty;

    - Recursivity: a workspace can include other workspace, which itself can include workspaces, etc. This allows notably to build incrementally complex services from simpler bricks;

    - Reusability: a workspace usually corresponds to a given functionality. This functionality can be more or less complex, according to the number of workspaces used recursively, and of other components. It can be used in the framework of

other services, thanks to the mechanism described above.

A workspace is defined by its .yaml configuration file.

- o Page: a Page is executed to provide the main front UI presented to the user;

- o Block: it is a graphical component which appears on a Page. It supports Block Protocol (https://blockprotocol.org/: an open source protocol; it allows notably interoperability and reusability). Example of Blocks: a webchat, a form, an interactive picture, etc.

- Backend:

  - Apps: a Workspace (as defined above) which has been published, and then usable by other Workspaces. The publication of a Workspace makes its current state is immutable. When an App is used in a Workspace, it is installed. By default, App features are public. It is of course possible to change this option to Private, as for any Object Oriented Programming Language;

  - Automations: they are either instructions, or triggers for these instructions. For instance (from Prisme Documentation: https://docs.eda.prisme.ai/en/workspaces/automations): an automation might send a notification message on Slack every time a new deal is created by sales teams.

### 3.1.2 Architecture

In Figure 1, we present the relations between the main framework components, as described above. The figure also displays the position of these components in functional terms.

Figure 2 below focuses on the interpretation of DSUL by the Runtime. This schema is a simplified version of the overall schema at https://docs.eda.prisme.ai/en/architecture. The overall schema displays other features, notably authentication, which we do not consider here: in this paper, we only focus on the possibilities afforded by the DSUL framework.

To summarize the process:

- The Runtime parses the .yaml file corresponding to a given workspace

- When a trigger is found, the Runtime execute it. An example of trigger is a call to an App or to an automation;

- The Runtime then sends the call to prisme.ai-events, through the Message Broker. The Message Broker "allows every internal microservice to communicate with each other" (Prisme documentation). Microservices are conceived so that they can be used on any Message Broker: it is an other improvement of reusability and robustness;

- If internal services are called, the communication is done directly through the Message Broker; the call to external services is done with API Gateway, through prisme.ai-events.

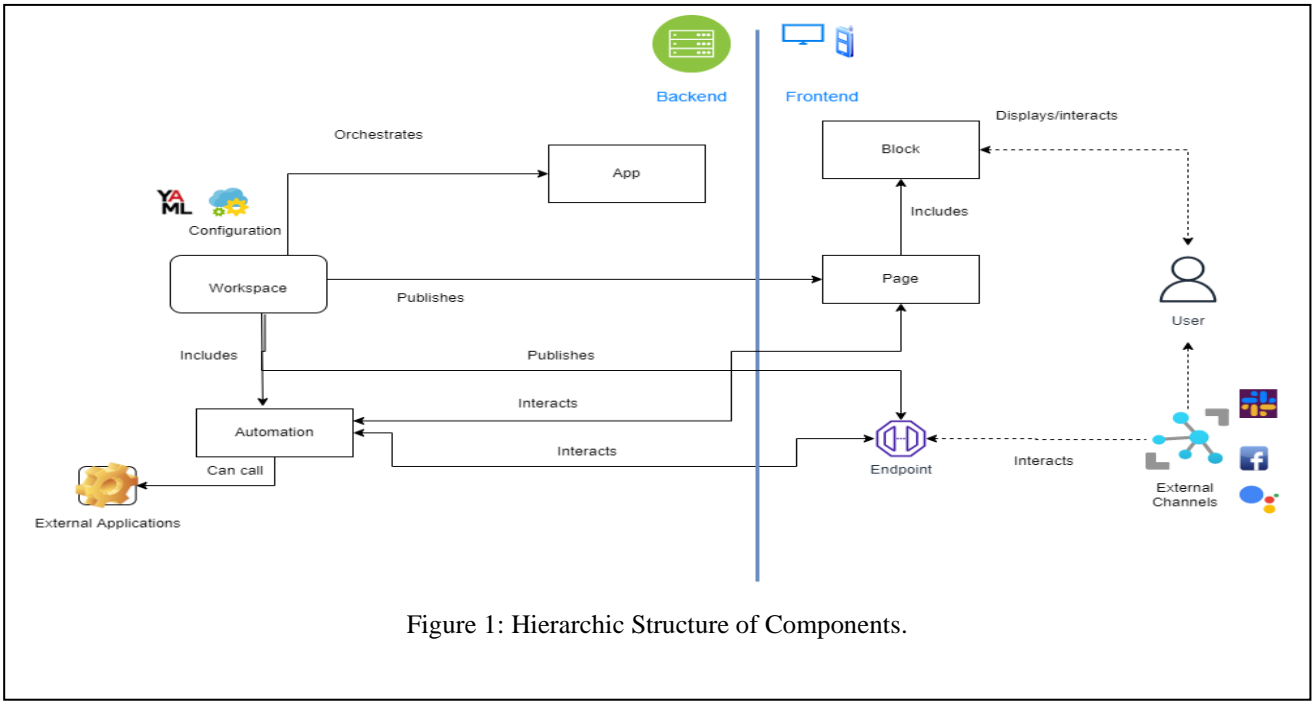

Figure 1: Hierarchic Structure of Components.

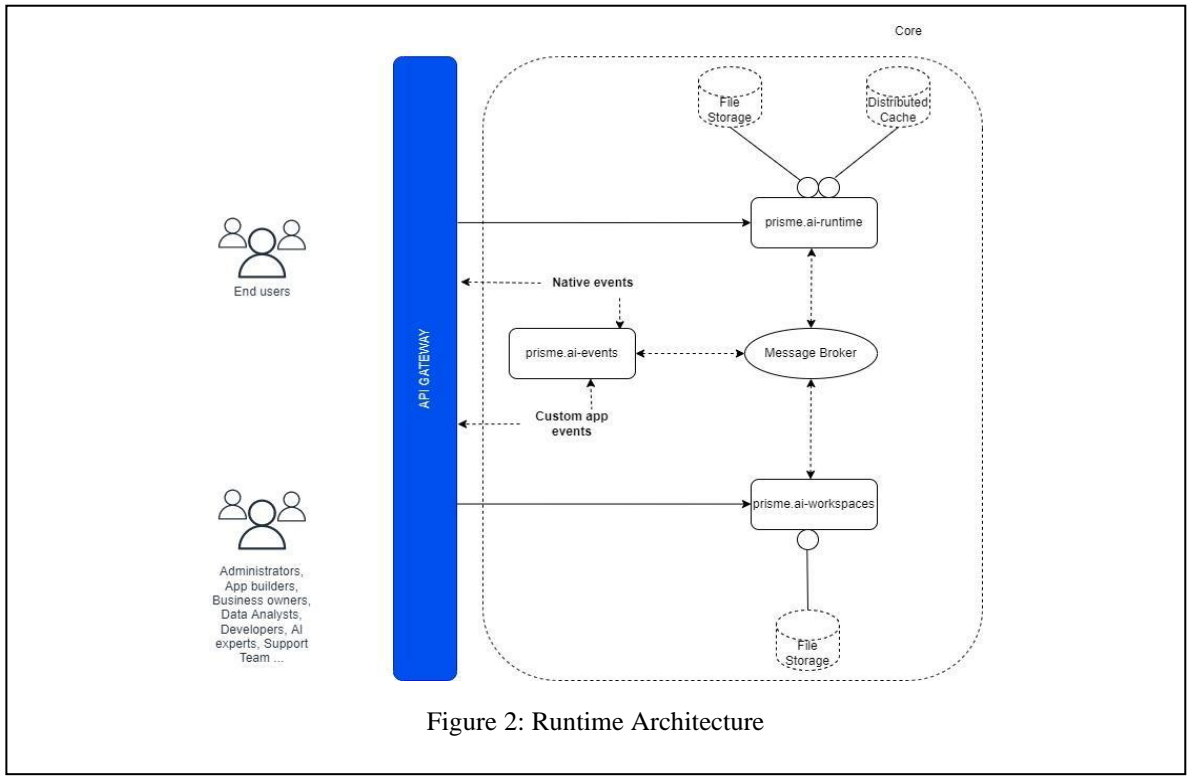

Figure 2: Runtime Architecture

In Figure 2, native events are handled by prisme.ai-runtime and by prisme.ai-workspaces; for instance: an automation has been added, an app has been installed, etc. Custom events, as for them, are specifically added by the developer using DSUL. Example: when an automation is called from a callbot, a custom event will handle the new call, with the audio file from the caller as a payload.

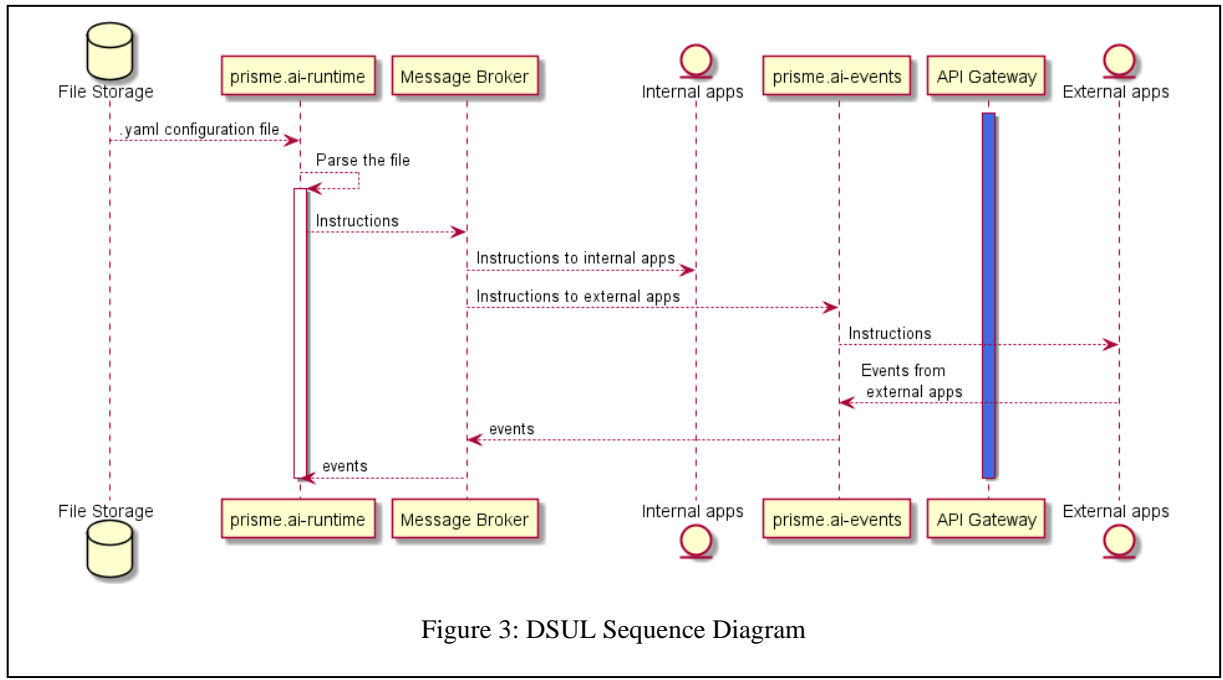

Figure 3: DSUL Sequence Diagram

The corresponding sequence diagram is described in Figure 3. It is simplified so that it displays only the processes relevant for this paper.

### 3.1.3 Examples

We present two examples of Digital services being created with DSUL. These examples are functionally and technically far from each other. Thus, we demonstrate that DSUL can be used for a wide array of various practical needs.
Let us remind that DSUL, besides these examples, can handle much more technical and functional variety.

### 3.1.3.. Example #1: a multicanal chatbot

The chatbot is specialized in booking a restaurant. The user can use either textual chat or vocal modality, both in Natural Language. When the booking is completed, a confirmation SMS is sent to the user's phone.
This solution uses several services:

- Vocal application: to handle Automatic Speech Recognition (ASR: vocal inputs) and Text To Speech (TTS: vocal outputs);

- Sending a SMS;

- Carry out the booking in a specialized application;

- Dialog Manager: to handle the dialog between the user and the system: intents detection, ask for confirmation, management of changes made by the user, etc.

- Language model trainer: Machine Learning module, used for Intent detection (Natural Language Understanding);

- Custom code: snippet of code for handling timeslot management.

These various services are coordinated in the .yaml file; it also handles their configuration.
DSUL allows the stacking of multiple software layers, from low level (basic functions) to higher level (calls to stacked functions). We illustrate it in the next example.

### 3.1.3.. Example #2: a multi-services Webpage

Here, we consider a Webpage allowing mail and OCR (Optical Character recognition) services.
This example illustrates the recursive use of Workspaces.
In the DSUL framework, this Webpage is a Workspace, which uses features provided by low-level software layers (NB: in this context, "low-level" corresponds to the level of embedment of a layer).

Let's consider the first layer of Workspace: it handles OCR function; the workspace uses an Automation to carry out the OCR.

This first Workspace is then published as an App.

This App is used by Workspace #2; this workspace allows adding OCR Function in a mail reader. Automations are created to add specific mail Features.

This second Workspace is then published as an App.

This App is used by Workspace #3; this workspace allows to add Mail Features to a Web page. Through the use of App/Workspace #2, it inherits the OCR and Mail Reader features.

This third Workspace also includes a Page, with Blocks. Thus, it eventually is usable as a Web Page with various functionalities.

### 3.2   Availability

The framework is specified and available at the url: `https://docs.eda.prisme.ai/api/`. It is distributed under Apache 2.0 license. The Runtime, as for it, is licensed under GPL V3.

## 4   Leads for Evaluation

In the first section, we describe the main features of DSUL which could be evaluated. In the second and last part, we propose a guideline for the comparison of our approach in regard to the main other.

### 4.1   Overview of evaluation protocols

In the literature, there are few papers evaluating low-code and alike approaches. We can cite (Calçada & Bernardino, 2021), who compare low-code with Java Swing and JavaScript frameworks. The standard defined by ISO 25010 (ISO 2014) for software quality can also be used to define metrics for evaluating our approach. There are also some papers, such as (Daniel *et al.*, op. cit.), which provides a solution and a method to evaluate it.
Numerous dimensions of coding can be submitted to evaluation. In this paper, we propose four criteria. Two relate to human factors: time of coding and ease of use. Two others are related to software metrics: runtime performance and green coding.

### 4.2   Evaluation for time saving

Saving of coding time can be ensured by several features. Among them are:

The learning curve to use a new software or framework: it can be measured by indicators provided by cognitive psychology, or by surveying developers (more details on this measure in next section);

- The volume of code needed to implement a function: usually the volume of code is measured in number of lines of code (LOC, cf. for instance (Rüdger *et al.*, 2008, p. 133 sq.) or (Honglei *et al.* 2009)). This measure is used by (Calcada & Bernadino, *op. cit.*, p. 105). Number of code lines may vary according to the way a program is written; though, some normalization is possible by referring to a standard code practice;

- The overall time necessary to code: this measure is also used by (Calcada & Bernadino, *op. cit.*, p. 105). The use of this measure is sensitive, since it may depend on other factors such as the coder's experiment and skill.

### 4.3   Evaluation for ease of use

Easiness of use is a subjective measure. Indeed, it is the coder/framework user who can state his opinion about this topic. To this end, a survey has to be designed to assess the targeted features for ease. For instance, (Daniel *et al.*, *op. cit.*, pp. 15432 sq.) submitted coders to 6 questions about their overall experience with the framework, the facility to use it, etc. To be significant enough, such surveys have to be submitted to several persons (20 users in (Daniel *et al.*, *op. cit.*)).

### 4.4   Evaluation for runtime performance

Runtime performance is not the main focus of our approach. However, this measurement is essential for any application aimed to end-users.

For a first evaluation, we ran tests on the code generated with DSUL. The detailed results of the tests are described at the url: `https://docs.eda.prisme.ai/en/architecture/technical/runtime/specifications/#performance-scalability`. The takeaway is that the run time for a user is around 20 ms. This demonstrates that there is no impact of DSUL overhead on the performances.

In the middle-term, we would position our platform performances in regard to the Computer Language Benchmarks Game (CLBG), which provide standard evaluations for 29 programming languages[3].

### 4.5 Evaluation for green coding

Green coding can be seen from two perspectives: code conception/writing and code runtime. The first perspective corresponds to the resources saved while writing and implementing the code. The second one addresses the resources consumed when the code is actually run. It is especially important for Web applications. Indeed, they can be displayed on a large number of terminals, and runs on multiple servers, at once; thus, they have a high impact for each hardware involved.

To our knowledge, the largest review of code energy efficiency is (Pereira *et al.* 2021). The authors evaluate the energy consumption of 27 programming languages, according to various criteria: memory, speed, etc. To do so, they use two resources as benchmarks: CLBG mentioned above, and Rosetta chrestomathy repository[4]. We will use this work as a baseline for our own evaluation of DSUL for energy consumption.

For code conception energy saving, a first basis are the metrics mentioned above related to time saving and overall volume of work required. These metrics should be used in combination with measures related to the resources used during the coding task. This last kind of measure has to be determined, notably in regard to the standard equipment used. The work mentioned above could also be useful to establish a baseline in this regard.

## 5 Comparison with other frameworks

In Table 1, we propose a quick overview of the main works on low code (academic and commercial). This selection is partial, since there are dozens of them; (Richardson & Rymer 2016, p. 2) list 42 vendors at the time of the report.

For generalization purposes, we categorize features from a macro perspective; for instance, "chatbots" category could itself subsume more precise features, as presented in section "Related Works" above.

| Features | DSUL | Appian | Salesforce | Mendix | Xatkit | AIML |
|---|---|---|---|---|---|---|
| Opensource | ✓ | | | | ✓ | ✓ |
| Machine Learning/AI | ✓ | ✓ | ✓ | ✓ | | |
| Chatbots | ✓ | | ✓ | | ✓ | ✓ |
| Web app | ✓ | ✓ | | ✓ | | |
| Mobile App | ✓ | | | ✓ | | |
| Processes | ✓ | ✓ | ✓ | | | |

Table 1: DSUL features in comparison with main related frameworks

---

[3] https://benchmarksgame-team.pages.debian.net/benchmarksgame/index.html

[4] https://rosettacode.org/wiki/Rosetta_Code

With this comparison, we see in a glance that DSUL covers more features than the other works, as stated in our description above.

## 6 Conclusion and Future Work

We presented DSUL, our framework for building omnichannel digital services. The innovation provided is the large cover for all possible applications, while remaining simple to use.

The state of art confirms that existing works, academic or industrial, are not as exhaustive. We then provide a thorough description of our system. We finally list the main leads for evaluating our system, from a human perspective but also for the sake of efficiency and performance.

Our next aim for DSUL is the direct generation of a service from one or several sentences in natural language. Thus, the user could obtain his digital services by only describing it in a short text. The same mechanism could allow to automatically generate services from other textual contents, such websites or specific posts in forums.

Another lead would be to extract the features of a given application, and represent them as a knowledge graph. This graph could be then used to generate a service in DSUL framework. One of the benefits would be to facilitate the benchmarking of an application generated by DSUL, in regard to a similar external service. More generally, we will proceed to an extensive evaluation of DSUL, following the leads we described.

## 7 References

AbuShawar, Bayan, & Atwell, Eric. 2015. "ALICE Chatbot: Trials and Outputs." *Computación y Sistemas*, 19(4), 625-632. https://doi.org/10.13053/CyS-19-4-2326

Arun S. Maiya. 2022. "ktrain: A Low-Code Library for Augmented Machine Learning", *Journal of Machine Learning Research 23* (2022) 1-6.

André Calçada and Jorge Bernardino. 2022. "Experimental Evaluation of Low Code development, Java Swing and JavaScript programming". In *International Database Engineered Applications Symposium (IDEAS'22)*, August 22–24, 2022, Budapest, Hungary. ACM, New York, NY, USA, 10 pages. https://doi.org/10.1145/3548785.3548792

G. Daniel, J. Cabot, L. Deruelle and M. Derras. 2020. "Xatkit: A Multimodal Low-Code Chatbot Development Framework," in *IEEE Access, vol. 8*, pp. 15332-15346, 2020, doi: 10.1109/ACCESS.2020.2966919.

Honglei, T., Wei, S. and Yanan, Z. (2009) "The Research on Software Metrics and Software Complexity Metrics". *International Forum on Computer Science-Technology and Applications*, Chongqing, 25-27 December 2009, Vol. 1, 131-136. https://doi.org/10.1109/IFCSTA.2009.39

ISO/IEC-25000:2014 Systems and software engineering – Systems and software Quality Requirements and Evaluation (SQuaRE). (2014). Retrieved from http://iso25000.com/index.php/en/iso-25000-standards/iso-25010?limit=3&start=6

Pereira R., Couto M., Ribeiro F., Rua R., Cunha J, Paulo Fernandes J, Saraiva J. 2021 "Ranking programming languages by energy efficiency", *Science of Computer Programming, Volume 205*, 2021, 102609, ISSN 0167-6423, https://doi.org/10.1016/j.scico.2021.102609

Richardson C. and Rymer J. 2016. "The Forrester Wave™: Low-Code Development Platforms", *Forrester Report*.

Richardson C. and Rymer J. 2014. "New Development Platforms Emerge For Customer-Facing Applications", *Forrester Report*.

Rüdiger Lincke, Jonas Lundberg, and Welf Löwe. 2008. "Comparing software metrics tools". *Proceedings of the 2008 international symposium on Software testing and analysis (ISSTA '08).* Association for Computing Machinery, New York, NY, USA, 131–142. https://doi.org/10.1145/1390630.1390648

M. S. Satu, M. H. Parvez and Shamim-Al-Mamun. 2015. "Review of integrated applications with AIML based chatbot," *International Conference on Computer and Information Engineering (ICCIE)*, 2015, pp. 87-90, doi: 10.1109/CCIE.2015.7399324

J. T. C. Tan and T. Inamura. 2012 "Extending chatterbot system into multimodal interaction framework with embodied contextual understanding" *7th ACM/IEEE International Conference on Human-Robot Interaction (HRI)*, 2012, pp. 251-252, doi: 10.1145/2157689.2157780